\documentclass[a4paper,11pt]{article}
\usepackage{pos}
\usepackage{graphicx}
\usepackage{caption}
\usepackage{subcaption}

\title{Implementation of machine learning techniques to predict impact parameter and transverse spherocity in heavy-ion collisions at the LHC}

\author*[a]{Aditya Nath Mishra}
\author[b]{Neelkamal Mallick }
\author[c]{Sushanta Tripathy}
\author[b]{Suman Deb}
\author[b,d]{Raghunath Sahoo}

\affiliation[a]{Wigner Research Centre for Physics\\
 29-33 Konkoly-Thege Mikl\'os str., 1121 Budapest, Hungary }

\affiliation[b]{Department of Physics, Indian Institute of Technology Indore\\
 Simrol, Indore 453552, India}
 
 \affiliation[c]{  INFN - sezione di Bologna\\
  via Irnerio 46, 40126 Bologna BO, Italy }
 
 \affiliation[d]{ CERN, CH 1211, Geneva 23, Switzerland }

\emailAdd{aditya.Nath.Mishra@wigner.hu}
\emailAdd{Neelkamal.Mallick@cern.ch}
\emailAdd{Sushanta.Tripathy@cern.ch}
\emailAdd{Suman.Deb@cern.ch}
\emailAdd{Raghunath.Sahoo@cern.ch}

\abstract{Machine learning techniques have been quite popular recently in the high-energy physics community and have led to numerous developments in this field. In heavy-ion collisions, one of the crucial observables, the impact parameter, plays an important role in the final-state particle production. This being extremely small (i.e. of the order of a few fermi), it is almost impossible to measure impact parameter in experiments. In this work, we implement the ML-based regression technique via Gradient Boosting Decision Trees (GBDT) to obtain a prediction of impact parameter in Pb-Pb collisions at $\sqrt{s_{NN}}$ = 5.02 TeV using A Multi-Phase Transport (AMPT) model. After its successful implementation in small collision systems, transverse spherocity, an event shape observable, holds an opportunity to reveal more about the particle production in heavy-ion collisions as well. In the absence of any experimental exploration in this direction at the LHC yet, we suggest an ML-based regression method to estimate centrality-wise transverse spherocity distributions in Pb-Pb collisions at $\sqrt{s_{NN}}$ = 5.02 TeV by training the model with minimum bias collision data. Throughout this work, we have used a few final state observables as the input to the ML-model, which could be easily made available from collision data. Our method seems to work quite well as we see a good agreement between the simulated true values and the predicted values from the ML-model.}

\FullConference{%
 
 The Ninth Annual Conference on Large Hadron Collider Physics - LHCP2021\\
 7-12 June 2021\\
 Online
}

\begin{document}
\maketitle

\section{Introduction}

The properties of hot and dense deconfined QCD matter, the Quark-Gluon Plasma (QGP), which is believed to be produced in high-energy heavy-ion collisions, are usually studied as a function of centrality classes of the collisions which are determined by the impact parameter ($b$)~\cite{Bass:1998vz}. However, obtaining the impact parameter values from experiments is still challenging as its value ranges in few femtometers ($fm$). Thus, in experiments, the centrality classes are inferred from final state charged-particle multiplicities and sometimes from the transverse energy distribution. In the hindsight, it would benefit the experiments if one can successfully implement Machine Learning (ML) based technique to obtain the impact parameter in a precise way from the final state observables.

ML techniques have enabled the development of tools that have played an important role in the field of high-energy physics (HEP) along with in different fields of physics for decades~\cite{Carleo:2019ptp,Ortiz:2020rwg}.  The impact parameter is one of the crucial physical quantities of heavy-ion collisions and it cannot be measured directly in experiments but might be inferred from observables at the final state.  ML algorithms such as support vector machine (SVM) and neural network (NN) have shown a great success in learning the complexity of large datasets for determination of the impact parameter in heavy-ion collisions~\cite{Bass:1993vx,David:1994qc,Bass:1996ez,Haddad:1996xw,DeSanctis:2009zzb,Li:2020qqn}.   
 In the present work, ML-based regression technique via GBDT is used to obtain predictions for impact parameter and spherocity distributions in Pb-Pb collisions at $\sqrt{s_{\rm NN}}$ = 2.76 and 5.02 TeV using A Multi-Phase Transport (AMPT)~\cite{AMPT2}. For ML, we have used a python based ML package, named as scikit-learn software package \cite{sklearn}. We have specifically used the {\em GradientBoostingRegressor} module inside {\em sklearn.ensemble} framework.
\begin{figure*}[ht!]
\includegraphics[scale=0.25]{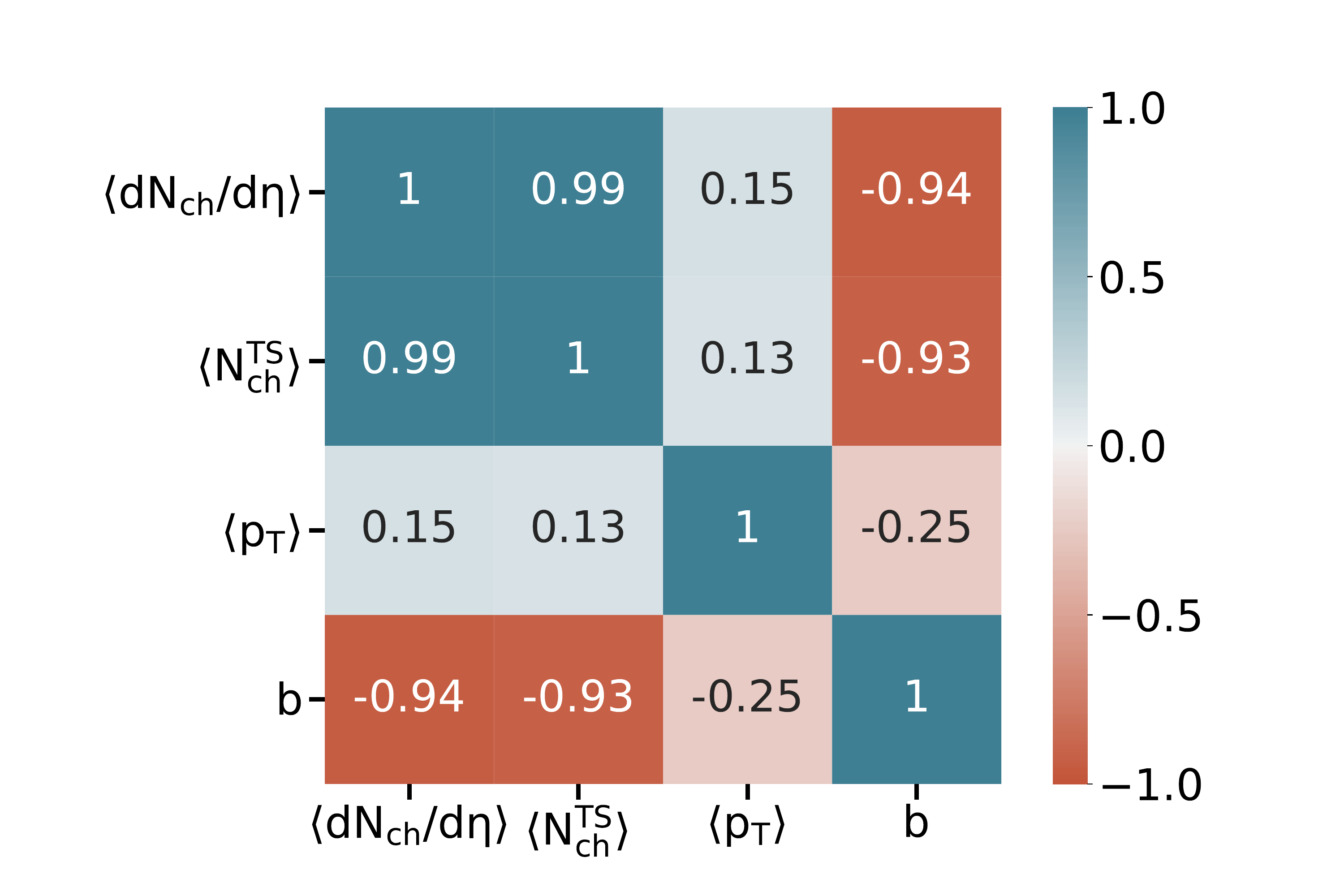}
\includegraphics[scale=0.25]{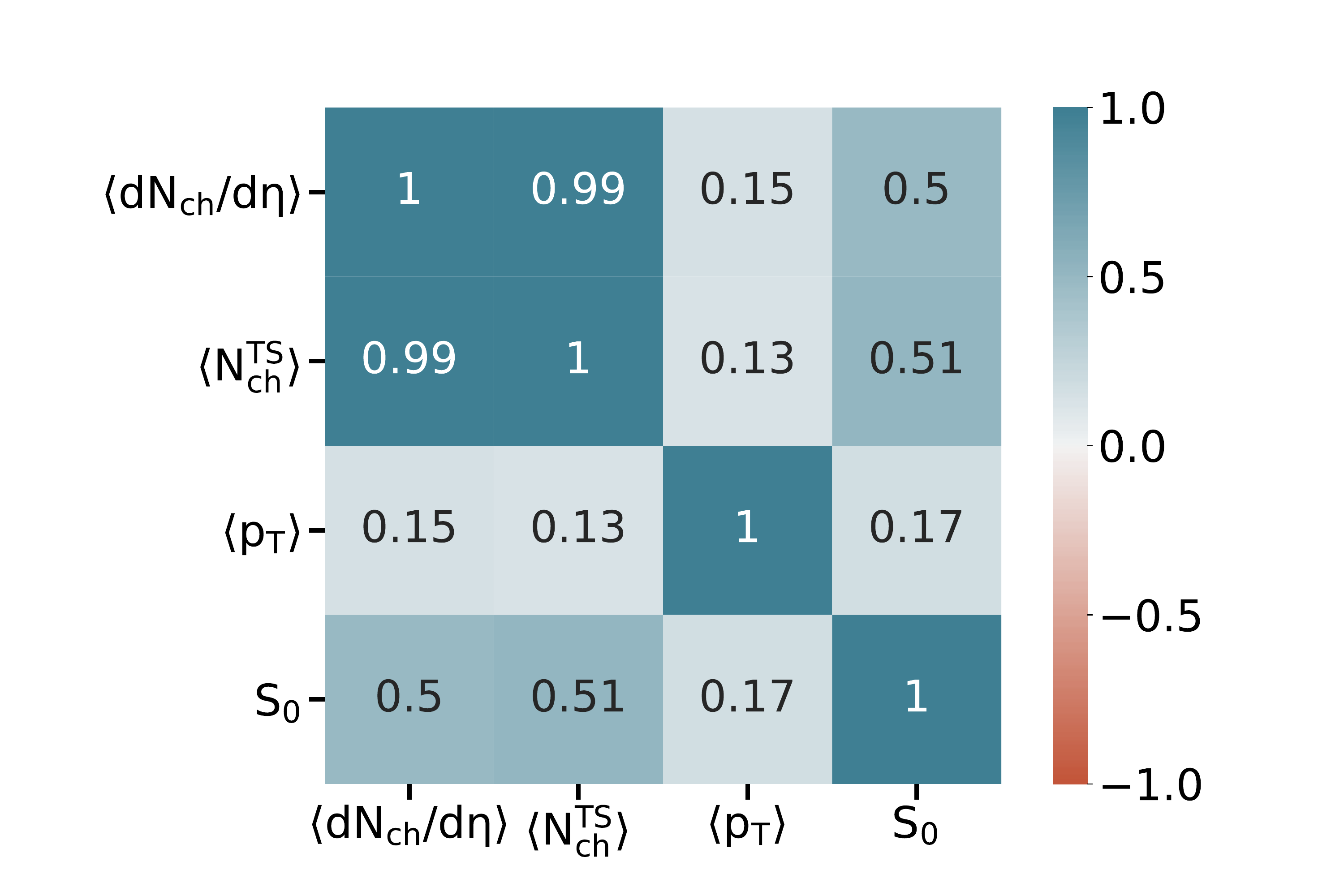}
\caption[]{(Color Online) Correlation matrix of the input variables ($\langle dN_{\rm ch}/d\eta \rangle$,   $\langle N_{\rm ch}^{\rm TS}\rangle$ and $\langle p_{\rm T} \rangle$ ) and target observables (impact parameter and transverse spherocity) in Pb-Pb collisions at $\sqrt{s_{\rm{NN}}} = 5.02$~TeV in AMPT model. The numbers show the correlation coefficients. The left panel shows the correlation matrix for impact parameter while the right panel shows the correlation matrix for transverse spherocity~\cite{Mallick:2021}.}
\label{CorMatrix}
\end{figure*}

\section{Results and Discussions}
\label{section4}

ML techniques are mainly used for classification, clustering and regression kind of problems. The problem addressed in this work is of supervised regression kind, {\em i.e.} for each set of the input variables, we have a finite numerical value as the target variable. Each set of the data refers to the final state observables of one event of the heavy-ion collisions. We have used charged-particle multiplicity ($\langle dN_{\rm ch}/d\eta \rangle$),  charged-particle multiplicity in the transverse region ( $\langle N_{\rm ch}^{\rm TS}\rangle$) and average transverse momentum ($\langle p_{\rm T} \rangle$) as the input variables and the target variables as the impact parameter ($b$) and transverse spherocity ($S_{0}$). For the problem discussed here, a widely used gradient boosting decision trees (GBDT) algorithm has been chosen.

Figure \ref{CorMatrix} represents the correlation matrix for the input variables and the target variables for Pb-Pb collisions at $\sqrt{s_{\rm NN}} = 5.02$ TeV minimum bias events. Here, left panel shows the correlation matrix for impact parameter while the right panel shows the correlation matrix for transverse spherocity. The numbers in the boxes represent the correlation coefficient which ranges from -1 to 1 and give the correlation strength between the intersecting variables in the matrix. The correlation coefficient ($\rho$) for two variables $x$ and $y$ is given by,
\begin{eqnarray}
\rho = \frac{{\rm cov}(x,y)}{\sigma_{x} \sigma_{y}}
\label{rho}
\end{eqnarray}
where ${\rm cov}(x,y)$ is the covariance and $\sigma_x$ and $\sigma_y$ are the standard deviations of $x$ and $y$ respectively. 

\begin{figure*}[h!]
\includegraphics[scale=0.45]{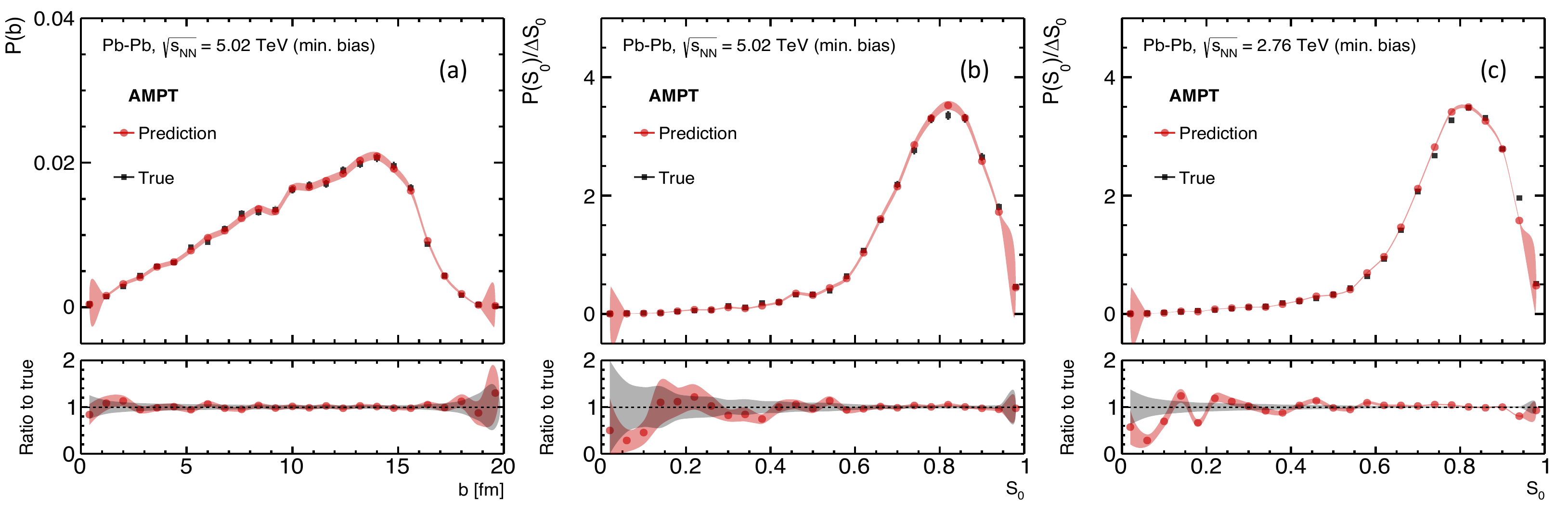}
\caption[]{(Color Online) Predictions for the impact parameter for Pb-Pb collisions at $\sqrt{s_{\rm{NN}}} = 5.02$~TeV and the transverse spherocity distributions for Pb-Pb collisions at $\sqrt{s_{\rm{NN}}} =$ 2.76 and 5.02~TeV using ML-based regression technique via GBDT in AMPT model are shown. The quadratic sum of the statistical and systematic uncertainties are shown as a red-colored band for the predicted values. The statistical uncertainties in the true values are shown as bars. In the ratio, black-colored band denotes the statistical uncertainties in the true values while the red-colored band denotes the quadratic sum of statistical and systematic uncertainties~\cite{Mallick:2021}.}
\label{Predictions-AMPT}
\end{figure*}

From the values of $\rho$, from the correlation matrix of the input variables and impact parameter, it is evident that there is a significant anti-correlation between impact parameter and the $\langle dN_{\rm ch}/d\eta \rangle$.  Also, impact parameter is found to be (anti-)correlated with the $\langle p_{\rm T} \rangle$ of an event. Figure~\ref{Predictions-AMPT}(a) shows the predictions for impact parameter distribution using ML for Pb-Pb collisions at $\sqrt{s_{\rm{NN}}} = 5.02$~TeV in AMPT model. The lower panel shows the ratio of predicted distribution to the true distribution. One can clearly see that the proposed ML framework with $\langle dN_{\rm ch}/d\eta \rangle$,   $\langle N_{\rm ch}^{\rm TS}\rangle$ and $\langle p_{\rm T} \rangle$ as the input variables, does a nice job of predicting the impact parameter distribution in Pb-Pb collisions at $\sqrt{s_{\rm{NN}}} = 5.02$~TeV.

 The values of correlation coefficients ($\rho$), from the correlation matrix of the input variables and transverse spherocity in Pb-Pb collisions at $\sqrt{s_{\rm{NN}}} = 5.02$~TeV, suggest that there is a sufficiently high correlation between the input variables ($\langle dN_{\rm ch}/d\eta \rangle$,   $\langle N_{\rm ch}^{\rm TS}\rangle$ and $\langle p_{\rm T} \rangle$) and the transverse spherocity of an event.
 Although, the correlation with $\langle p_{\rm T} \rangle$  is small but it is still significant for a proper prediction of transverse spherocity through ML-based regression technique via GBDT.  Based on the mentioned inputs, predictions for transverse spherocity distribution in Pb-Pb collisions at $\sqrt{s_{\rm{NN}}} = 5.02$~TeV is shown in Fig.~\ref{Predictions-AMPT}(b). Here. the predicted spherocity distribution is compared with the true ones obtained from AMPT. One can clearly see that the proposed ML-based regression technique via GBDT predicts the spherocity distribution accurately in Pb-Pb collisions at $\sqrt{s_{\rm{NN}}} = 5.02$~TeV. However, at low-spherocity regions, we see a deviation from the true distribution and this could be due to the fact that in heavy-ion collisions the statistics of having events with theback-to-back structure are expected to be quite less compared to events with isotropic in nature. Thus, we believe that this deviation could be due to limited statistics in the low spherocity region, which can also be seen by the statistical uncertainty black-colored band in the lower panel. In Fig.~\ref{Predictions-AMPT}(c). We have also successfully predicted the  spherocity distribution for Pb-Pb collisions at $\sqrt{s_{\rm{NN}}} = 2.76$~TeV using the ML training from Pb-Pb collisions at $\sqrt{s_{\rm{NN}}} = 5.02$~TeV in wide spherocity ranges. This suggests that the correlation of spherocity distributions with the input variables are quite similar across LHC energies. For more details about the methodology used in this article, one can see Ref.~\cite{Mallick:2021}.

\section{Summary}
\label{section5}
The impact parameter and transverse spherocity are two key observables in high-energy heavy-ion collisions. In this work, we implement the ML-based regression technique via Gradient Boosting Decision Trees to obtain a prediction of impact parameter and transverse spherocity in the midrapidity minimum bais Pb-Pb collisions at the LHC energies using A Multi-Phase Transport Model (AMPT) model. For this study, we use final state charged-particle multiplicity and mean transverse momentum as the input variables for the training of the ML algorithm. 

In the absence of experimental measurements, we propose to implement ML-based regression technique to obtain transverse spherocity from the observed final state quantities in heavy-ion collisions. We would like to mention here that the ML-based training with the correlations of input observables using a MC model is quite useful when the MC model describes the input observables as close as possible to the experimental data. This method will be useful to handle the physics associated with unmeasured quantities in the experiment.

\section*{Acknowledgements}
R.S. acknowledges the financial supports under the CERN Scientific Associateship and the financial grants under DAE-BRNS Project No. 58/14/29/2019-BRNS. The authors would like to acknowledge the usage of resources of the LHC grid computing facility at VECC, Kolkata. S.T. acknowledges the support from INFN postdoctoral fellowship in experimental physics.  A.N.M.  would like to thank the Hungarian National Research, Development and Innovation Office (NKFIH) under the contract numbers OTKA K135515, K123815 and NKFIH 2019-2.1.11-TET-2019-00078, 2019-2.1.11-TET-2019-00050, Wigner GPU Laboratory,

\end{document}